\documentclass[
    ,final            
  ]
  {aipproc}

\layoutstyle{6x9}

\begin{document}

\title{Large-x PDFs and the Drell-Yan Process}

\classification{13.85.Qk, 14.20.Dh, 24.85.+p, 13.88.+e} 

\keywords      {Drell-Yan, quarkonium production, parton distributions}

\author{J.C.~Peng}{
  address={University of Illinois at Urbana-Champaign, Urbana, IL 61801}
}

\begin{abstract}
Dimuon production has been studied in a series of fixed-target
experiments at Fermilab during the last two decades. Highlights
from these experiments, together with recent
results from the Fermilab E866 experiment, are presented. Future
prospects for studying the parton distributions in the nucleons
and nuclei using dimuon production are also discussed.
\end{abstract}

\maketitle


\section{Introduction}

The Drell-Yan process~\cite{drell70}, in which a charged lepton pair
is produced in a hadron-hadron interaction via the
electromagnetic $q \bar q \to l^+ l^-$ process, has provided
unique information on parton distributions. In particular,
the Drell-Yan process has been used to determine the antiquark
contents of nucleons and nuclei~\cite{pat99}, as well as
the quark distributions of pions, kaons, and antiprotons~\cite{kenyon82}.
Such information is difficult, if not impossible, to obtain from
DIS experiments. As the Drell-Yan process can be well
described by next-to-leading order QCD calculations~\cite{stirling93},
a firm theoretical framework exists for utilizing the
Drell-Yan process to extract the parton distributions.

During the last two decades, a series of fixed-target dimuon
production experiments (E772, E789, E866) have been carried out
using 800 GeV/c proton beam at Fermilab.
At 800 GeV/c, the dimuon data contain Drell-Yan continuum up to
dimuon mass of $\sim 15$ GeV as well as quarkonium productions
(J/$\Psi$, $\Psi^\prime$, and $\Upsilon$ resonances). The Drell-Yan
process and quarkonium productions often provide complementary
information, since Drell-Yan is an electromagnetic process
via quark-antiquark annihilation while the quarkonium production
is a strong interaction process dominated by gluon-gluon fusion
at this beam energy.

The Fermilab dimuon experiments covers a broad range of physics
topics. The Drell-Yan data have provided informations on the
antiquark distributions in the nucleons~\cite{pat92,hawker98,
peng98,towell01} and nuclei~\cite{alde90,vasiliev99}.
These results showed the surprising results that the
antiquark distributions in the
nuclei are not enhanced~\cite{alde90,vasiliev99},
contrary to the predictions of models which explain the
EMC effect in term of nuclear
enhancement of exchanged mesons. Moreover, the Drell-Yan cross
section ratios $p+d/p+p$ clearly establish the
flavor asymmetry of the $\bar d$ and $\bar u$ distributions in
the proton, and they map out the $x$-dependence of this
asymmetry~\cite{hawker98,peng98,towell01}.
Pronounced nuclear dependences of quarkonium productions have
been observed for J/$\Psi$, $\Psi^\prime$, and $\Upsilon$ resonances
~\cite{alde91a,alde91b,kowitt94,leitch95}.

Several review articles covering some of these results are
available~\cite{pat99,garvey01,reimer07}. In this article, we
will focus on the recent results from experiment E866 and
future prospect of dimuon experiments at Fermilab and
J-PARC.

\section{FLAVOR STRUCTURE OF LIGHT-QUARK SEA}

In the CERN NA51~\cite{na51a} and Fermilab E866~\cite{hawker98,
peng98,towell01}
experiments on proton-induced dimuon production, a striking difference was
observed for the Drell-Yan cross sections between $p+p$ and $p+d$.
As the underlying mechanism for the Drell-Yan process involves
quark-antiquark annihilation, this difference has been attributed to
the asymmetry between the up and down sea quark distributions
in the proton. From the $\sigma (p+d)_{DY}/2\sigma(p+p)_{DY}$ ratios
the Bjorken-$x$ dependence of the sea-quark
$\bar d / \bar u$ flavor
asymmetry has been extracted~\cite{hawker98,peng98,towell01,na51a}.

\begin{figure}[tb]
\includegraphics*[width=\linewidth]{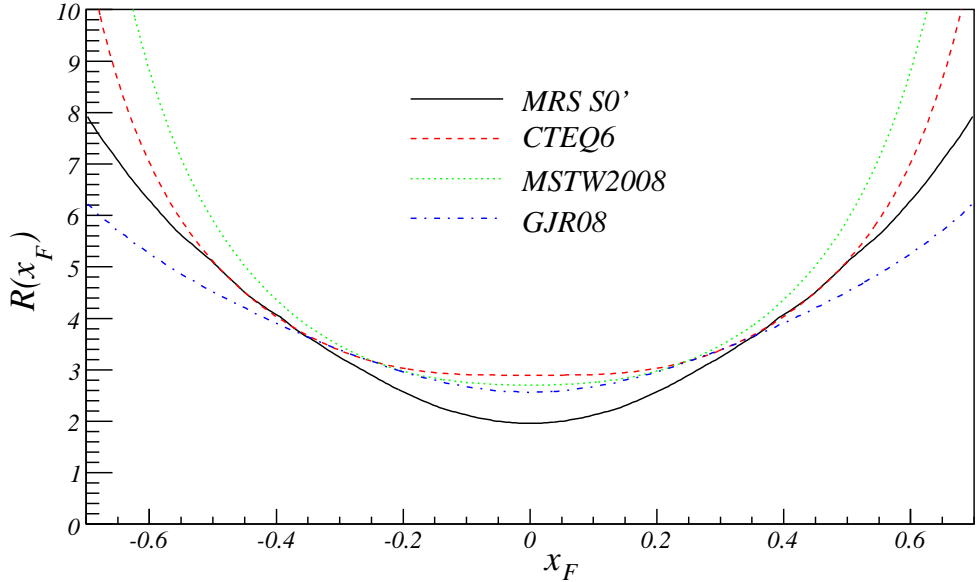}
\caption{Prediction of the ratio $(p+p \to W^+ +x)/(p+p \to W^- +x)$ at $\sqrt s
$ of 500 GeV using various PDFs.}
\label{crosfig1}
\end{figure}

Future fixed-target dimuon experiments have been proposed at
the 120 GeV Fermilab Main Injector (FMI) and the 50 GeV J-PARC
facilities. The Fermilab proposal~\cite{e906}, E906, has been
approved and is expected to start data-taking around 2011. Two
dimuon proposals (P04~\cite{p04} and P24~\cite{p24}) have also
been submitted to the J-PARC for approval. The lower beam energies
at FMI and J-PARC present opportunities for extending the $\bar d/\bar u$
and the nuclear antiquark distribution measurements to larger
$x$ ($x>0.25$). For given values of $x_1$ and $x_2$, the Drell-Yan cross
section is proportional to $1/s$, hence a gain of $\sim 16$ times in the
Drell-Yan cross sections can be obtained at the J-PARC energy of 50 GeV.
Since the perturbative process gives a symmetric $\bar d/ \bar u$ while
non-perturbative processes are necessary to generate an asymmetric
$\bar d/ \bar u$ sea, it would be very important to extend the Drell-Yan
measurements to kinematic regimes beyond the current limits. Another
advantage of lower beam energies is that a much more sensitive study of the
partonic energy loss in nuclei could be carried out using the Drell-Yan
nuclear dependence~\cite{garvey03}.

To disentangle the $\bar d / \bar u$ asymmetry from the possible
charge-symmetry violation effect~\cite{ma1,boros},
one could consider $W$ boson production in $p + p$ collision at RHIC.
An interesting quantity to be measured is the ratio of the
$p + p \to W^+ + x$ and $p + p \to W^- + x$ cross sections~\cite{peng1}.
It can be shown that this
ratio is very sensitive to $\bar d / \bar u$. An important feature of
the $W$ production asymmetry in $p + p$ collision
is that it is completely free
from the assumption of charge symmetry.
Another advantage of $W$ production in $p+p$ collision is that no
nuclear effects need to be considered. Finally, the $W$ production is
sensitive to $\bar d/ \bar u$ flavor
asymmetry at a $Q^2$ scale of $\sim$ 6500 GeV$^2/$c$^2$, significantly
larger than all existing measurements. This offers the opportunity to
examine the QCD evolution of the sea-quark flavor asymmetry.
Figure 1 shows the
predictions~\cite{yang09} for $p + p$ collision at $\sqrt s =
500~$GeV. The MRS S0$^\prime$ corresponds to the $\bar
d/\bar u$ symmetric parton distributions, while the other three
parton distribution functions are from recent global fits with asymmetric
$\bar d/\bar u$ sea-quark distributions. Figure 1 clearly shows that
$W$ asymmetry
measurements at RHIC could provide an independent determination
of $\bar d / \bar u$. First results from the LHC are also expected
soon, and will provide additional information~\cite{yang09}.

Unlike the electromagnetic
Drell-Yan process, quarkonium
production is a strong interaction dominated by the subprocess of
gluon-gluon fusion at 800 GeV beam energy. Therefore, the
quarkonium production cross
sections are primarily sensitive to the gluon distributions in the
colliding hadrons. The $\Upsilon$ production ratio, $\sigma(p+d \to \Upsilon)
/2 \sigma(p+p \to \Upsilon)$, is expected to
probe the gluon content in the neutron relative to that in the
proton~\cite{piller}.

\begin{figure}[tb]
\includegraphics*[width=\linewidth]{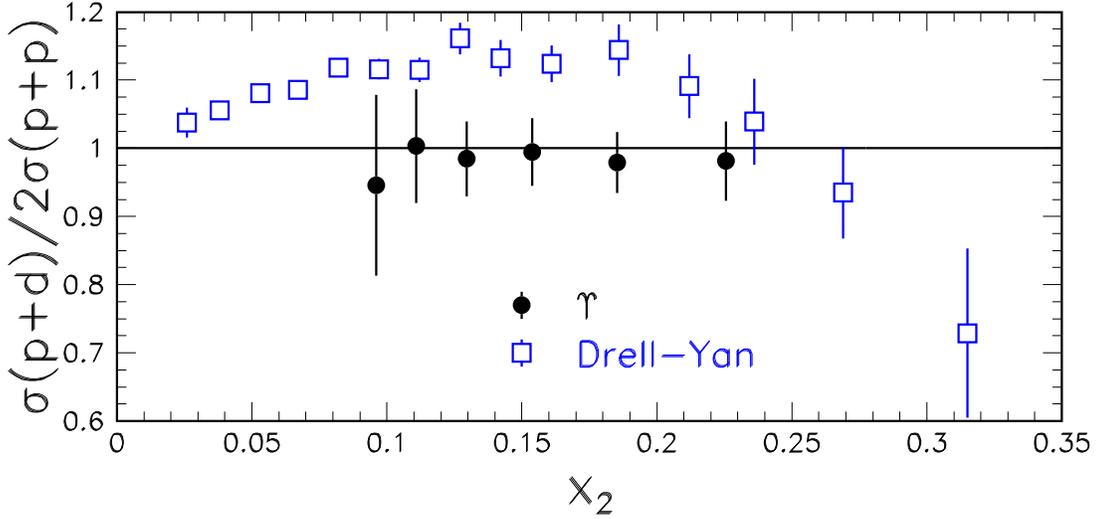}
\caption{The E866 $\sigma(p + d) /2 \sigma(p + p)$ cross section
ratios for $\Upsilon$ resonances as a function of $x_2$. The
corresponding ratios for Drell-Yan cross sections
are also shown. The error bars are statistical only.}
\label{crosfig2}
\end{figure}

The $\sigma (p+d)/2\sigma(p+p)$ ratios for $\Upsilon (1S+2S+3S)$ production
are shown in Fig.\  2 as a function of $x_2$~\cite{zhu08}.
These ratios are consistent with unity, in striking
contrast to the corresponding values for the Drell-Yan process,
also shown in Fig.\  2.
The difference between the Drell-Yan and the
$\Upsilon$ cross section ratios clearly reflect the different
underlying mechanisms in these two processes.
For $\Upsilon$ production, the dominance of the
gluon-gluon fusion subprocess at this beam energy
implies that $\sigma (p+d \to \Upsilon)/ 2 \sigma(p+p \to \Upsilon)
\approx \frac {1}{2} (1+g_n(x_2)/g_p(x_2))$. Figure 2 shows that
the gluon distributions in the proton ($g_p$) and neutron ($g_n$)
are very similar over the $x_2$ range $0.09 < x_2 < 0.25$. The overall
$\sigma (p+d \to \Upsilon)/ 2 \sigma(p+p \to \Upsilon)$ ratio, integrated
over the measured kinematic range, is $0.984 \pm 0.026 (\rm{stat.})
\pm 0.01 (\rm{syst.})$.
The $\Upsilon$ data indicate that the gluon distributions in the
proton and neutron are very similar.

The upcoming Fermilab E906 experiment is expected to provide a precise
measurement of the $\sigma (p+d \to J/\Psi)/ 2 \sigma(p+p \to J/\Psi)$
ratio. These data should further test the equality of the gluon
distributions in the proton and neutron. Moreover, these data could
identify EMC effects for deuteron. It is interesting to note that
analogous ratios measured in DIS and the Drell-Yan can not be used
to determine the EMC effects for deuteron, since different parton
distributions are being probed for the hydrogen and the deuterium targets.
If one assumes that the gluon distributions are identical in proton and
neutron, then any deviation of the 
$\sigma (p+d \to J/\Psi)/ 2 \sigma(p+p \to J/\Psi)$ ratio from unity
would be attributed to the EMC effects of gluon distributions in deuteron.

\section{Flavor Dependence of the EMC Effect}

Despite a quarter century of significant experimental and theoretical effort,
the specific origins of the observed $A$ dependence of the nuclear quark
distributions have yet to be unambiguously identified. Attempts to explain the
EMC effect have led to a large collection of theoretical
models~\cite{geesaman_review,norton_review}, many of which are
capable of describing the essential features of the data,
however the underlying
physics mechanisms in each model are often very different.

A new calculation of the modifications of nucleon quark
distributions in the nuclear medium has recently been
reported~\cite{ianemc,ian}.
In this approach by Clo\"{e}t, Bentz and Thomas (CBT), the
Nambu--Jona-Lasinio model is used to describe the coupling of the quarks
in the bound nucleons to the scalar and vector mean fields inside a nucleus.
These nucleon quark distributions are then
convoluted with a nucleon momentum distribution in the nucleus to generate the
nuclear quark distributions~\cite{ianemc}. An important feature of this model
is that for $N\neq Z$ nuclei (where $N$ and $Z$ refer to the number of
neutrons and protons) the isovector--vector mean field (usually denoted
by $\rho^0$) will affect the up quarks differently from the down quarks
in the bound nucleons. Therefore, this model has a novel prediction that
the $u$ and $d$ quarks have distinct nuclear modifications for
$N\neq Z$ nuclei.

Semi-inclusive DIS (SIDIS) on heavy nuclear targets, in which the flavor 
of the struck quark is tagged by the detected hadron, is a
promising experimental tool to search for the flavor-dependent EMC effect.
Recently, Lu and Ma~\cite{ma2006} pointed out that charged lepton 
SIDIS off nuclear targets and the deuteron can be used to probe the 
flavor content of the nuclear quark sea, which can help distinguish 
between the various models of the EMC effect. Indeed, a SIDIS 
experiment~\cite{pr12004} aiming at a precise determination of flavor 
dependence of the EMC effect has also been proposed at the
upgraded $12\,$GeV JLab facility.

\begin{figure}[tb]
\includegraphics*[width=\linewidth]{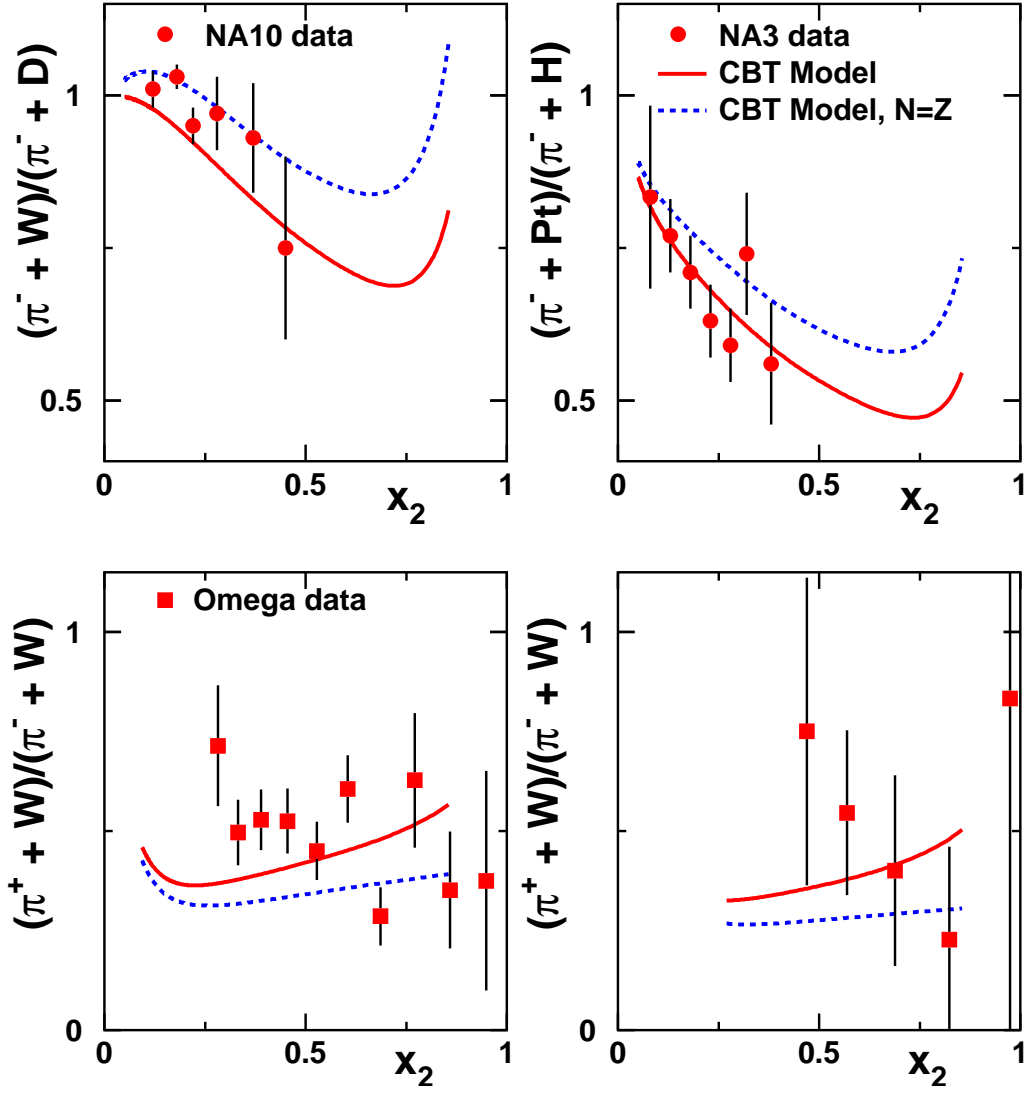}
\caption{The existing data for the ratios
$\frac{\sigma^{DY}(\pi^- + W)}
{\sigma^{DY}(\pi^- + D)}$ (upper left),
$\frac{\sigma^{DY}(\pi^- + Pt)}
{\sigma^{DY}(\pi^- + H)}$ (upper right) and
$\frac{\sigma^{DY}(\pi^+ +W)}
{\sigma^{DY}(\pi^- + W)}$ (lower) versus the predictions
using nuclear PDFs
of the CBT model for tungsten (red solid) and $N=Z$ nuclear matter
(blue dashed).}
\label{crosfig3}
\end{figure}

The flavor dependence of the EMC effect is a
promising experimental observable to distinguish
among the plethora of models that can describe the EMC effect.
As discussed earlier, the nuclear dependence of proton-induced 
Drell-Yan process measured at Fermilab has shown the surprising 
results that the antiquark distributions in the
nuclei are not enhanced~\cite{alde90,vasiliev99}, contrary to the
predictions of many EMC models.
Pion-induced Drell-Yan process provides another
experimental tool with which search for flavor-dependent effects
in the nuclear modification of the nucleon structure functions~\cite{dutta2010}.

To explore the sensitivity of pion-induced Drell-Yan processes to
a flavor-dependent EMC effect, we consider the three ratios
$\frac{\sigma^{DY}(\pi^+ + A)}{\sigma^{DY}(\pi^- + A)}$, 
$\frac{\sigma^{DY}(\pi^- + A)} 
{\sigma^{DY}(\pi^- + D)}$ and
$\frac{\sigma^{DY}(\pi^- + A)}{\sigma^{DY}(\pi^- + H)}$,
where $A$ represents a nuclear, D a deuteron and H a hydrogen target.
Assuming isospin symmetry, which implies
$u_{\pi^+} = d_{\pi^-}$, $\bar{u}_{\pi^-} = \bar{d}_{\pi^+}$,
$\bar{u}_{\pi^+} = \bar{d}_{\pi^-}$, $u_{\pi^-} = d_{\pi^+}$ and
keeping only the dominant terms in each cross-section, one readily obtains

\begin{equation}
R_{\pm} = \frac{\sigma^{DY}(\pi^+ + A)}{\sigma^{DY}(\pi^- + A)}
\approx \frac{d_{A}(x)}{4u_{A}(x)}
\end{equation}
\begin{equation}
R^{-}_{A/D} = \frac{\sigma^{DY}(\pi^- + A)}{\sigma^{DY}(\pi^- + D)}
\approx \frac{u_{A}(x)}{u_{D}(x)}
\end{equation}
\begin{equation}
R^{-}_{A/H} = \frac{\sigma^{DY}(\pi^- + A)}{\sigma^{DY}(\pi^- + H)}
\approx \frac{u_{A}(x)}{u_p(x)},
\end{equation}

\noindent The up and down nuclear quark distributions are labeled by $u_A$ and
$d_A$ respectively, $u_{D}$ is the up quark distribution in the deuteron
and $u_p$ the up quark distribution in the proton.
Eqs. 1-3 demonstrate
that these Drell-Yan cross-section ratios are very sensitive to the
flavor dependence of the EMC effect.

Fig. 3 shows the comparison~\cite{dutta2010} of 
the calculations of the pion-induced
Drell-Yan cross-section ratios with the existing data. The top
left panel shows the ratio of
$\frac{\sigma^{DY}(\pi^- + W)}{\sigma^{DY}(\pi^- + D)}$
from the NA10 experiment~\cite{na10}. These plots contain both the
$P_{beam} = 286$ and $140\,$GeV data sets, which are very similar.
The calculations are performed at $P_{beam} = 286\,$GeV, since most of
the data was obtained at this energy.
The PDFs of the CBT model~\cite{ianemc,ian} at a fixed $Q^2$ of $25\,$GeV
$^2$, which is approximately the mean $Q^2$ of the NA10 experiment,
were used. The top right panel shows the ratio
$\frac{\sigma^{DY}(\pi^- + Pt)}{\sigma^{DY}(\pi^- + H)}$ from
the NA3 experiment~\cite{na3}. The data were collected using a
$150\,$GeV $\pi^-$ beam and the $Q^2$ range covered was
$16.8 \leq Q^2\leq70.6\,$GeV$^2$. The calculations are performed for
$P_{beam} = 150\,$GeV and $Q^2 = 25\,$GeV$^2$.

The solid curves in Fig. 3 are calculations using the flavor-dependent
$u_A(x)$ and $d_A(x)$ from the CBT model with $N/Z=1.5$, corresponding
approximately to the $N/Z$ values for the Au, W and Pt nuclei.
The dashed curves correspond to the calculated ratios using the nuclear
PDFs from the CBT model for $N=Z$. Since $u_A/u_D = d_A/d_D$ in this case,
the dashed curves are representative of the predictions for flavor-independent
EMC models. Figure 3 shows that the NA10 data do not exhibit a clear
preference for the flavor-dependent versus flavor-independent nuclear PDFs.
In contrast, the NA3 data strongly favor the calculations using
flavor-dependent nuclear PDFs. The existing
Drell-Yan data are not sufficiently accurate yet, although the NA3 data
clearly favor the flavor-dependent over the flavor-independent nuclear PDFs.
Precise future pion-induced Drell-Yan experiments can provide
unique constraints that will help distinguish the various theoretical models
and most importantly shed new light on the origins of the EMC effect.

\section{Transverse Spin and Drell-Yan Process}

The study of the transverse momentum dependent (TMD) parton distributions
of the nucleon has received much attention in recent years
as it provides new perspectives on the hadron structure and
QCD~\cite{barone02}. These novel TMDs can be extracted from semi-inclusive
deep-inelastic scattering (SIDIS) experiments. Recent measurements of
the SIDIS by the HERMES~\cite{hermes05}
and COMPASS~\cite{compass05} collaborations
have shown clear evidence for the existence of the T-odd Sivers functions.
These data also allow the first determination~\cite{vogelsang05} of
the magnitude and flavor structure of the Sivers functions and the nucleon
transversity distributions.

The TMD and transversity parton distributions can also be probed in
Drell-Yan experiments. As pointed out~\cite{ralston79}
long time ago, the double
transverse spin asymmetry in polarized Drell-Yan, $A_{TT}$, is proportional
to the product of transversity distributions, $h_1(x_q)h_1(x_{\bar q})$.
The single transverse spin asymmetry, $A_N$, is sensitive to the
Sivers function~\cite{sivers90}, $f^\perp_{1T}(x)$ of
the polarized proton (beam or target).
Even unpolarized Drell-Yan experiments can be used to probe the TMD
distribution function, since the cos$2\phi$ azimuthal angular dependence
is proportional to the product of two Boer-Mulders functions~\cite{boer98},
$h^\perp_1(x_1) \bar h^\perp_1(x_2)$. A unique feature of the Drell-Yan
process is that, unlike the SIDIS, no fragmentation functions are involved.
Therefore, the Drell-Yan process provides an entirely independent technique
for measuring the TMD functions. Furthermore, the proton-induced Drell-Yan
process is sensitive to the sea-quark TMDs and can lead to flavor separation
of TMDs when combined with the SIDIS data. Finally, the intriguing
prediction~\cite{collins02} that the T-odd TMDs extracted from
DIS will have a sign-change for the
Drell-Yan process remains to be tested experimentally.

No polarized Drell-Yan experiments have yet been performed.
However, some information on the
Boer-Mulders functions have been extracted recently
from the azimuthal angular distributions in the unpolarized Drell-Yan
process. The general expression for the Drell-Yan
angular distribution is~\cite{lam78}
\begin{equation}
\frac {d\sigma} {d\Omega} \propto 1+\lambda \cos^2\theta +\mu \sin2\theta
\cos \phi + \frac {\nu}{2} \sin^2\theta \cos 2\phi,
\label{eq:eq1}
\end{equation}
\noindent where $\theta$ and $\phi$ are the polar and azimuthal decay angle
of the $l^+$ in the dilepton rest frame. Boer showed that the $\cos 2\phi$
term is proportional to the convolution of the quark and antiquark
Boer-Mulders functions in the projectile and target~\cite{boer99}.
This can be understood by noting that the Drell-Yan cross
section depends on the transverse spins of the annihilating quark and
antiquark. Therefore, a correlation between the transverse spin and
the transverse momentum of the quark, as represented by the Boer-Mulders
function, would lead to a
preferred transverse momentum direction.

Pronounced $\cos 2 \phi$ dependences
were indeed observed in the NA10~\cite{falciano86} and E615~\cite{conway89}
pion-induced Drell-Yan experiments, and attributed to the
Boer-Mulders function.
The first measurement of the $\cos 2 \phi$
dependence of the proton-induced Drell-Yan process was recently reported for
$p+p$ and $p+d$ interactions
at 800 GeV/c~\cite{zhu07}. In contrast to pion-induced Drell-Yan,
significantly smaller (but non-zero) cos$2\phi$ azimuthal angular dependence
was observed in the $p+p$ and $p+d$ reactions. While the 
pion-induced Drell-Yan process
is dominated by annihilation between a valence antiquark in the pion
and a valence quark in the nucleon, the
proton-induced Drell-Yan process involves a valence quark in the proton
annihilating with a sea antiquark in the nucleon. Therefore, the
$p+p$ and $p+d$ results suggest~\cite{zhu07,zhang08} that 
the Boer-Mulders functions for
sea antiquarks are significantly smaller than those for valence quarks.

\section{Future Prospects}

Future fixed-target dimuon experiments have been proposed at
the 120 GeV Fermilab Main Injector and the 50 GeV J-PARC
facilities. As discussed earlier, the Fermilab E906 experiment
will extend the $\bar d/ \bar u$ asymmetry measurement to
larger $x$ region. Another goal of this experiment is to determine
the antiquark distributions in nuclei at large $x$ using nuclear
targets. New information on the quark energy loss in nuclei is
also expected. As discussed earlier, an
advantage of lower beam energies is that a much more sensitive study of the
partonic energy loss in nuclei could be carried out~\cite{garvey03}.

With the possibility to accelerate polarized proton beams 
at J-PARC~\cite{p24},
the spin structure of the proton can also be investigated with the
proposed dimuon experiments. In particular, polarized Drell-Yan process
with polarized beam and/or polarized target at J-PARC would allow a unique
program on spin physics complementary to polarized DIS experiments
and the RHIC-Spin programs. Specific physics topics include the measurements
of T-odd Boer-Mulders distribution function in unpolarized Drell-Yan,
the extraction of T-odd Sivers distribution functions in singly
transversely polarized Drell-Yan, the helicity distribution of
antiqaurks in doubly longitudinally polarized Drell-Yan, and the
transversity distribution in doubly transversely polarized
Drell-Yan. It is worth
noting that polarized Drell-Yan is one of the major physics program at the
GSI Polarized Antiproton Experiment (PAX). The COMPASS experiment at CERN
will also measure $\pi^-$-induced Drell-Yan on transversely polarized
targets to extract Sivers functions from the Single-Spin-Asymmetry.
Two recent Letter-of-Intent have also been submitted to RHIC for measuring
polarized Drell-Yan at RHIC-Spin, one in collider mode and the other
utilizing internal targets. It is clear that the Drell-Yan process will
continue to provide unique information on parton distributions in the
forseeable future.

\bibliographystyle{aipproc}

\end{document}